\documentclass[reqno,12pt]{article}
\usepackage{amsmath,amsfonts,amssymb,amsthm,amstext,amscd,eucal}

\usepackage{graphicx}

\usepackage{slashed,mathrsfs,cite}
\usepackage[all]{xy}
\newcommand{\mathsym}[1]{{}} 
\usepackage[colorlinks=true,
            linkcolor=blue,
            urlcolor=blue,
            citecolor=blue]{hyperref}
\usepackage{enumerate}
\DeclareMathAlphabet{\pazocal}{OMS}{zplm}{m}{n}
\usepackage{graphicx}
\usepackage{bm}

\usepackage{hyperref}
\usepackage{color,enumerate}

\definecolor{DeepPink2}{rgb}{0.93,0.07,0.54}
\definecolor{purple4}{rgb}{0.33,0.10,0.55}

\makeatletter \@addtoreset{equation}{section}

\makeatletter\renewcommand\section{\@startsection {section}{1}{\z@}%
                                   {-3.5ex \@plus -1ex \@minus -.2ex}
                                   {2.3ex \@plus.2ex}%
                                   {\normalfont\large\bfseries}}
\renewcommand\subsection{\@startsection{subsection}{2}{\z@}%
                                     {-3.25ex\@plus -1ex \@minus -.2ex}%
                                     {1.5ex \@plus .2ex}%
                                     {\normalfont\bfseries}}
\DeclareMathAlphabet{\mathcal}{OMS}{cmsy}{b}{n}

\newcommand{\orcid}[1]{\href{https://orcid.org/#1}{\includegraphics[width=10pt]{orcid}}}

\makeatother

\newcommand{\email}[1]{\footnote{E-mail: \href{mailto:#1}{#1}}}

\begin{document}

\title{\bf\Large{ On the photon mass generation in Rarita-Schwinger QED}}

\author{\textbf{M.~Ghasemkhani \email{m$_{_{-}}$ghasemkhani@sbu.ac.ir } $^{a}$, G.~Soleimani  \email{g.soleimani.m76@gmail.com} $^{a}$, A. Soto \email{alex.soto@newcastle.ac.uk} $^{b}$  and R.~Bufalo \email{rodrigo.bufalo@ufla.br}  $^{c}$}\\\\
\textit{\small$^{a}$Department of Physics, Shahid Beheshti University, 1983969411, Tehran, Iran}\\
\textit{\small$^{b}$School of Mathematics, Statistics and Physics, Newcastle University,}\\
\textit{\small	Newcastle upon Tyne, NE1 7RU, UK}\\
\textit{\small$^{c}$Departamento de F\'isica, Universidade Federal de Lavras,}\\
\textit{\small Caixa Postal 3037, 37200-900 Lavras, MG, Brazil}\\
}

\maketitle
\date{}

\begin{abstract}
This work examines the dynamical mass generation for the photon in Rarita-Schwinger QED.
We focus our attention on the cases of $\omega=2,3$ dimensional spacetime.
In these frameworks, it is well known that in the usual QED, the photon field (dynamically) acquires a gauge invariant mass (the Schwinger and Chern-Simons mass, respectively).
We wish to scrutinize this phenomenon in terms of the Rarita-Schwinger fields.
The presence of higher-derivative terms is shown as the leading contributions to the $1$PI function $\langle AA \rangle$ at one-loop order.
We study the pole structure of the photon's complete propagator to unveil the main effects of the Rarita-Schwinger fields on the photon's mass.
In addition, we present some remarks about the renormalizability of this model (in different dimensions) due to the presence of higher-derivative corrections at one-loop.
\end{abstract}

\setcounter{footnote}{0}
\renewcommand{\baselinestretch}{1.05}  

\newpage

\section{Introduction}

Higher-spin field theories (with spin content $\geq 3/2$) have been extensively studied over the years in different contexts \cite{Bengtsson:2020,Bonora:2016ida,Bonora:2016otz,Bonora:2017ykb}, mainly in self-consistent constructions of grand unification theories \cite{Adler:2014pga}, by improving the UV behavior of Einstein's gravity \cite{Campoleoni:2024ced} as well as the natural emergence of these theories in the AdS/CFT correspondence \cite{Giombi:2016ejx}, among others.
Although free propagation of fields of arbitrary spins is allowed, the structure of interactions is strongly spin-dependent \cite{Singh:1974qz,Singh:1974rc,Fronsdal:1978rb,Fang:1978wz}. Actually,
introducing interactions leads to consistency problems, e.g. it does not preserve the physical degrees of freedom of the higher-spin particle.

Since the proposal of the Rarita-Schwinger (RS) theory in the description of the spin-$3/2$ field \cite{Rarita:1941mf}, it has found great phenomenological interest in many contexts, such as the gravitinos of supergravity \cite{Freedman:1976py,Das:1976ct,Gates:1983nr}, scattering of spin-$3/2$ particles \cite{Delgado-Acosta:2009ulg}, modelling hadron resonances \cite{deJong:1992wm,Pascalutsa:1999zz,Bernard:2003xf}, and also in Lorentz violating scenarios  \cite{Gomes:2022btc,Gomes:2023qkj}, among others.
Despite possessing interesting physical behavior and being phenomenologically significant, the RS theory presents certain drawbacks which are constantly scrutinized in order to cast it in a consistent form \cite{Moldauer:1956zz,Johnson:1960vt,Velo:1969txo,Velo:1969bt,Aurilia:1969bg,Nath:1971wp,Adler:2015yha,Adler:2015zha}.

It is well-known that low-dimensional models have a plethora of interesting aspects, ranging from applications in condensed matter physics
to their use as toy models.
Although there is some interest in higher-spin field theories in three-dimensional spacetime, for instance, higher spin/CFT correspondence on weakly coupled conformal field theories (CFT) \cite{Prokushkin:1998bq,Maldacena:2011jn,Maldacena:2012sf}, there are features that have not been deeply studied.
Actually, when we focus on Rarita-Schwinger fields, the majority (if not the totality) of the studies are in four-dimensional models; hence, it would be interesting to study Rarita-Schwinger fields as matter content of low-dimensional models.

Undoubtedly, the most prominent property of 3-dimensional fermionic models is that the fermion mass parameter $m$ breaks parity invariance and this feature has nontrivial consequences even when $m \to 0$.
In this sense, we wish to further analyze the behavior of gauge fields coupled with RS fields in this case.\footnote{Actually, in three dimensions, the little group does not admit massless representations of arbitrary helicity, so that only scalar and spin-$1/2$ degrees of freedom can propagate \cite{Binegar:1981gv}. Nonetheless, besides examining topological properties in 3D, one can still consider these field theories as toy models for higher-spin theories in 4D.}
This study seeks to explore the robust symmetry content of the RS field in well-known aspects of the gauge field.

Therefore, we will address the question of whether the dynamical mass generation for the photon in $\omega=2,3$ is manifest in the presence of RS fields minimally coupled with an Abelian gauge field.
To discuss these aspects, we start a review in Sec.~\ref{sec2}  about the definition of the Lagrangian density for the RS-QED in a general $\omega$-dimensional spacetime, where a discussion about the free propagator for the RS field and its interaction with a gauge field is presented.
In Sec.~\ref{sec3}, we compute explicitly the $1$PI function at one-loop order (polarization tensor) in $\omega=2,3,4$, where we observe the presence of higher-derivative terms as leading contributions.
We give some remarks about the renormalizability of the model in the presence of these terms.
In Sec.~\ref{sec4}, we study the pole structure of the photon's complete propagator to examine the non-trivial poles $p^2 \neq 0$, in which we note that the higher-derivative terms change completely the propagator poles in comparison with the usual QED.
At last, we present our final remarks and prospects in Sec.~\ref{sec5}.

\section{Abelian Rarita-Schwinger QED}
\label{sec2}

In this section, we review the main aspects regarding the definition of the Rarita-Schwinger field coupled with an electromagnetic field in a general $\omega$-dimensional space-time.

Our starting point is to establish the Lagrangian density for free spin-$\frac{3}{2}$ fields in $\omega$ dimensions \cite{Pilling:2004cu}:
\begin{align} \label{eq1}
\mathscr{L} =& \, \bar{\psi}_{\mu} \bigg[ \left( i \slashed{\partial} - m \right) \eta^{\mu
\nu} - i \frac{1}{\omega - 1} \left( 1 - \frac{\omega - 2}{\omega} a \right) (\gamma^{\mu}
\partial^{\nu} + \partial^{\mu} \gamma^{\nu})  \cr
&  + \frac{1}{\omega - 1} \left( 1 +
\frac{\omega - 2}{\omega^2} a^2 \right) i \gamma^{\mu} \slashed{\partial} \gamma^{\nu} + \frac{1}{\omega - 1} \left( 1 + \frac{(2 + a)}{\omega} a \right) m \gamma^{\mu}
\gamma^{\nu} \bigg] \psi_{\nu},
\end{align}
in which $a$ is a real free parameter of the theory \footnote{As a matter of fact, this parameter is related to the point transformation of the RS fields \cite{Pilling:2004cu,Freedman:1976py}}.

The $\omega$-dimensional form of the Lagrangian \eqref{eq1} is obtained by imposing a series of conditions upon the operator $\Lambda^{\mu\nu}$ defined in the generalized form as $\mathscr{L} =  \bar{\psi}_{\mu}\Lambda^{\mu\nu}   \psi_{\nu} $ \cite{Aurilia:1969bg,Moldauer:1956zz,Nath:1971wp,Pilling:2004cu},  described below:

\begin{enumerate}[(i)]
\item  The field equation should be the same famous Rarita-Schwinger equation for a spin-$\frac{3}{2}$ field;
\item  The Lagrangian should be non-singular in the limit $p\to 0$,

\item  The Lagrangian should be linear in derivatives (fermionic fields),
\item  The respective action should be hermitian.
\end{enumerate}

Moreover, the expression \eqref{eq1} can be cast into the usual Rarita-Schwinger form when $a=-\omega$ \cite{Pilling:2004cu,Bernard:2003xf, deJong:1992wm, Pascalutsa:1999zz}, so that
\begin{equation} \label{eq2}
\mathscr{L} = \bar{\psi}_{\mu} \left[ \left( i \slashed{\partial} - m \right) \eta^{\mu
\nu} - i (\gamma^{\mu} \partial^{\nu} + \partial^{\mu} \gamma^{\nu}) + i
\gamma^{\mu} \slashed{\partial} \gamma^{\nu} + m \gamma^{\mu} \gamma^{\nu} \right]
\psi_{\nu}.
\end{equation}
This expression is used to determine the free propagator for the massive RS field \footnote{Moreover, we shall also consider the propagator for the massless RS field in sec.~\ref{schwinger}, so that we can discuss the generation of the Schwinger mass in $\omega=2$.}
\begin{equation} \label{eq4}
S^{\mu \nu} (p) = \frac{\slashed{p} + m}{p^2 - m^2} \left[ \eta^{\mu \nu} -
\left(\frac{1}{\omega - 1}\right) \gamma^{\mu} \gamma^{\nu} - \left(\frac{1}{\omega - 1}\right) \frac{\gamma^{\mu}
p^{\nu} - p^{\mu} \gamma^{\nu}}{m} -\left( \frac{\omega - 2}{\omega - 1}\right) \frac{p^{\mu}
p^{\nu}}{m^2} \right].
\end{equation}

On the other hand, the interaction between the electromagnetic and RS fields is introduced in terms of the minimal coupling, it means promoting the partial derivatives to covariant ones $\partial_{\mu} \rightarrow D_{\mu} =
\partial_{\mu} - i g A_{\mu}$ \cite{Pilling:2004cu,Gomes:2022btc},  which yields
\begin{equation} \label{eq3}
\mathscr{L} = \bar{\psi}_{\mu} \left[ \left( i \slashed{D} - m \right) \eta^{\mu \nu} - i
(\gamma^{\mu} D^{\nu} + D^{\mu} \gamma^{\nu}) + i \gamma^{\mu} \slashed{D}
\gamma^{\nu} + m \gamma^{\mu} \gamma^{\nu} \right] \psi_{\nu},
\end{equation}
where the RS field transforms as the usual matter so that the model is gauge invariant under the local $U(1)$ transformations $A_\mu \to A_\mu -\partial_\mu \chi $ and $\psi_\mu \to e^{i\chi } \psi_\mu $.
In this case, the vertex part of \eqref{eq3} reads as
\begin{equation} \label{eq5}
\mathscr{L}_{\rm int} = g \bar{\psi}_{\mu}  \left[\gamma^{\rho} \eta^{\mu \nu} - (\gamma^{\mu} \eta^{\nu
\rho} + \eta^{\mu \rho} \gamma^{\nu}) + \gamma^{\mu} \gamma^{\rho}
\gamma^{\nu}\right] A_{\rho} \psi_{\nu}.
\end{equation}
At last, the Feynman vertex rule can be read off from  \eqref{eq5} and it results into
\begin{equation} \label{eq6}
\Gamma^{\mu \rho \nu} = i g \left(\gamma^{\mu} \gamma^{\rho} \gamma^{\nu} +
\gamma^{\rho} \eta^{\mu \nu} - \gamma^{\mu} \eta^{\nu \rho} - \eta^{\mu \rho}\gamma^{\nu}\right).
\end{equation}

Since our main interest is to examine the effects of RS fields on the (dynamically generated) mass of the gauge field in $\omega=2,3$, the quantity that we shall study is the photon polarization tensor.
Furthermore, once the interaction is due to the cubic vertex $\langle \bar{\psi} A \psi \rangle $, we have the contribution of one graph as depicted in Fig.~1.
Hence, the Feynman amplitude for the polarization tensor in $\omega$ dimensions is given by
\begin{equation} \label{eq7}
i\Pi^{\mu \nu} (p) = - \int \frac{d^{\omega} k}{(2 \pi)^{\omega}}~{\rm tr} \Big[\Gamma^{\alpha
\mu\sigma} i S_{\sigma \xi} (k) \Gamma^{\xi \nu \lambda} i S_{\lambda\alpha}(k + p)\Big].
\end{equation}

Our next step is to evaluate explicitly this expression for $\omega=2,3,4$, in which we shall observe the presence of gauge invariant parity even (Maxwell term) and also parity odd  (Chern-Simons term) contributions, when the spacetime dimensionality is odd.

\begin{figure}[t]
	\includegraphics[height=3.4\baselineskip]{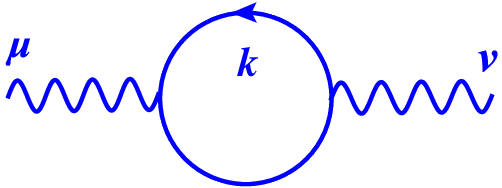}
	\centering\caption{One-loop photon polarization graph.}
	\label{fig:oneloop}
\end{figure}

\section{Two-point function for the Rarita-Schwinger model}
\label{sec3}

Although the expression for the polarization tensor \eqref{eq7} is more complicated than the usual QED (due to the propagator \eqref{eq4} and the vertex \eqref{eq6}), it can be evaluated using the dimensional regularization with help of the FeynCalc Mathematica package \cite{Mertig:1990an,Shtabovenko:2016sxi,Shtabovenko:2020gxv,Shtabovenko:2023idz}.
Hence, the parity even sector of the two-point function amplitude \eqref{eq7} in $\omega$-dimensions is
\begin{align} \label{eq8}
i\Pi^{\mu\nu}(p)&= -\frac{ i g^2}{m^4} ~ \Gamma\left(2-\frac{\omega}{2}\right)
\left(p^2 \eta^{\mu\nu}-p^{\mu}p^{\nu}\right) \int_{0}^{1} dx~\frac{ 2^{(1-\omega)} }{\pi^{\frac{\omega}{2}} (\omega-1)^2}\big(m^2+p^2\left(x-1\right)x\big)^{\frac{\omega-4}{2}} \nonumber\\
&\times \bigg\{
 p^2 x\left(x-1\right)\Big[\big(\omega\left((\omega-4)\omega+2\right)+8\big)m^2+ \left(\omega+1\right)\left(\omega-2\right)^2  \left(x-1\right)x p^2 \Big]
\nonumber\\
&+m^4\Big[\omega(2-\omega)\big((\omega-1)^2(x-1)x+2\big)
+4\Big]\bigg\},
\end{align}
which clearly satisfies the Ward identity $p_{\mu}\Pi^{\mu\nu}(p)=p_{\nu}\Pi^{\mu\nu}(p)=0$.
This expression will be studied now in various cases, for $\omega=2,3,4$, discussing some aspects related to their differences with the usual QED counterpart.
Moreover, the parity odd sector of \eqref{eq7} will be evaluated below because it will be examined directly for $\omega = 3$.

\subsection{The case for $\omega=2$ }
\label{schwinger}

The amplitude of the photon self-energy \eqref{eq8}  for $\omega=2$ is solved exactly and it has a finite result as shown below
\begin{equation} \label{PT_W2}
	i\Pi^{\mu\nu}(p)=-\frac{2ig^2}{\pi m^2}\left(p^2\eta^{\mu\nu}-p^{\mu}p^{\nu}\right).
\end{equation}
This result is very similar to those obtained in the usual massive Schwinger model \cite{Coleman:1975pw}, where the photon does not acquire a massive pole (in contrast with the massless Schwinger model).
We shall examine some features of this model below.

On the other hand, to discuss the dynamical mass generation for the photon (Schwinger mass) we must consider massless RS fields.
We observe that for the case $m=0$, the operator $\Lambda^{\mu\nu}$ is transverse, which shows that the theory displays a gauge symmetry ($\delta \bar{\psi}_\mu = \partial_\mu \Theta$, where $\Theta$ is an arbitrary spinor).
Hence, it is necessary to introduce a gauge fixing term for the RS field so that we can define the propagator consistently.
Finally, in a covariant gauge, the propagator for massless RS fields is
\cite{Manoukian:2016rxf,Lindwasser:2023zwo}
\begin{equation} \label{prop_mass}
\widetilde{S}^{\mu\nu}\left(p\right)=\frac{1}{p^{2}-i\epsilon}\left(\eta^{\mu\nu}\slashed{p}+\frac{\gamma^{\mu} \slashed{p} \gamma^{\nu}}{\omega-2}\right).
\end{equation}
We observe an apparent singular point at $\omega \to 2$ in the propagator \eqref{prop_mass}.
In this case, the polarization tensor \eqref{eq7} can be evaluated using dimensional regularization and it reads
\begin{equation}
\widetilde{\Pi}^{\mu\nu}\left(p\right)=ig^{2}\omega\left(\omega-1\right)\Gamma\left(2-\frac{\omega}{2}\right)\frac{\left(p^{2}\eta^{\mu\nu}-p^{\mu}p^{\nu}\right)}{p^{2}}\int_{0}^{1}dx~\frac{2^{1-\omega}}{\pi^{\frac{\omega}{2}}}\left[p^{2}x\left(x-1\right)\right]^{\frac{\omega-2}{2}}.
\end{equation}
Moreover, the limit $\omega \to 2$ can be readily evaluated and yields a finite and exact result as follows
\begin{equation} \label{PT_W2_mass}
\widetilde{\Pi}^{\mu\nu}\left(p\right)=\frac{ig^{2}}{\pi p^{2}}\left(p^{2}\eta^{\mu\nu}-p^{\mu}p^{\nu}\right).
\end{equation}
This expression has a pole at $p^2=0$, displaying a similar structure of the massless QED$_2$, so we expect to observe the photon's mass generation. We address this point below.

\subsection{The case for $\mathbf{\omega}=3$}

To evaluate the one-loop photon's amplitude for the case of $\omega=3$, we should be careful to consider both parity even and odd sectors from expression \eqref{eq7} (this can be also done with help of the FeynCalc Mathematica package \cite{Mertig:1990an,Shtabovenko:2016sxi,Shtabovenko:2020gxv,Shtabovenko:2023idz}).
After some algebraic manipulations involving the traces of $\gamma$ matrices (recall that for  $\omega=3$ we have ${\rm tr} \left( \gamma^{\mu} \gamma^{\nu} \gamma^{\alpha} \right) = 2i \epsilon ^{\mu\nu \alpha }$) and computing the momentum integration, we obtain
\begin{align} \label{eq11}
i\Pi^{\mu\nu}(p) &=	\frac{i g^{2}}{48\pi m^{4}}\int_{0}^{1} dx~ \frac{1}{\sqrt{m^{2}+p^{2} x (x-1)}} \cr
& \times \biggl\{ 2\left[p^{2} x(x-1)\left(4 p^{2} x(x-1) +5m^{2}\right)-2m^{4}\big(1+6x(x-1)\big)\right]\left(p^{\mu}p^{\nu}-p^{2}\eta^{\mu\nu}\right)\cr
&+3 \left[m^{2} p^{2} \big(1- 16x(x-1)\big)+2p^{4}x (x-1)-12m^{4}\right] im\epsilon^{\mu\nu\alpha}p_{\alpha}\biggr\}.
\end{align}

This expression cannot be solved exactly, thus, to gain insight about its physical behavior we shall consider the low-energy limit, $p^2\ll m^2$.
Hence, for the parity-even part of \eqref{eq11}, we have
\begin{equation} \label{eq10}
i\Pi^{\mu\nu}(p)\Big|_{\rm P-even}= \frac{ig^2}{30\pi m^3}~p^2 \left[p^2\eta^{\mu\nu}-p^{\mu}p^{\nu}\right]-\frac{ig^2}{480\pi m^5}~p^4 \left[p^2\eta^{\mu\nu}-p^{\mu}p^{\nu}\right]
+{\cal{O}}(m^{-7}).
\end{equation}
The first point to remark in \eqref{eq10} is that the usual Maxwell term ${\cal{O}}(m^{-1})$ is not (dynamically) generated in the
 $\rm RS$-$\rm QED_3$; this is a significant change in comparison with the $\rm QED_3$.
Actually, the first non-vanishing term in \eqref{eq10} is the higher-derivative (HD) Maxwell term ${\cal{O}}(m^{-3})$.
Hence, we observe that the effective electromagnetic theory generated in $\rm RS$-$\rm QED_3$ is completely due to the HD terms.
Moreover, since these induced HD terms are finite they do not render problems regarding renormalization.

On the other hand, the parity-odd sector of \eqref{eq11} reads
\begin{equation} \label{eq10b}
i\Pi^{\mu\nu}(p)\Big|_{\rm P-odd}=\frac{3g^2}{4\pi}\epsilon^{\mu\nu\alpha}p_{\alpha}-\frac{g^2}{6\pi m^2}~p^{2}\epsilon^{\mu\nu\alpha}p_{\alpha}+ {\cal{O}}(m^{-4}).
\end{equation}
Here, similar to $\rm QED_3$, we see that the ordinary Chern-Simons term ${\cal{O}}(m^{0})$ is induced and corrected by HD contributions.
It is worth mentioning that we shall have a significant change in the propagating modes for the photon in $\rm RS$-$\rm QED_3$, when compared with the $\rm QED_3$, because while the parity-odd sector of the effective action starts with the usual CS term, the parity-even sector starts with the HD Maxwell term.
More details about this point are given below.

By means of completeness, we establish a comparison among $\rm RS$-$\rm QED_3$, spinor $\rm QED_3$ and scalar $\rm QED_3$, by presenting a summary of their main results about the induced one-loop effective action generated in these models in the table \ref{tab1}.
As we observe, the ordinary Chern-Simons and HD Chern-Simons actions are generated in spinor $\rm QED_3$ and $\rm RS$-$\rm QED_3$ at order ${\cal{O}}(m^{-2\ell})$   while these actions are not induced in scalar $\rm QED_3$ at all.
Also, the ordinary Maxwell and HD Maxwell actions are produced in spinor $\rm QED_3$ and scalar $\rm QED_3$  at order ${\cal{O}}(m^{-(2\ell+1)})$ while only
HD Maxwell action is present in $\rm RS$-$\rm QED_3$ (initiated with $\ell=1$).

This table shows the explicit dependence of the photon's effective action on the spin of coupled (integrated) matter.

\begin{table}[h!]
\centering
	\begin{tabular}{|c|c|c|c|c|}
		\hline
		\color{blue}{Order of expansion} &  \color{blue}{Induced action} & \color{blue}{Spinor $\rm QED_3$} & \color{blue}{ Scalar $\rm QED_3$}& \color{blue}{$\rm RS$-$\rm QED_3$} \\
		\hline
		\hspace{-0.2cm}\color{blue}{${\cal{O}}(m^{0})$} & Chern-Simons & {\rm Yes} & {\rm No} & {\rm Yes} \\
		\hline
		\color{blue}{${\cal{O}}(m^{-1})$} & Maxwell & {\rm Yes} &  {\rm Yes} & {\rm No} \\
		\hline
		\color{blue}{${\cal{O}}(m^{-2})$} & HD Chern-Simons & {\rm Yes} & {\rm No} & {\rm Yes}\\
		\hline
		\color{blue}{${\cal{O}}(m^{-3})$} & HD Maxwell & {\rm Yes} & {\rm Yes} & {\rm Yes} \\
		\hline
		\vdots & \vdots & \vdots & \vdots & \vdots \\
		\hline
		\color{blue}{${\cal{O}}(m^{-2\ell})$} & HD Chern-Simons & {\rm Yes} & {\rm No} & {\rm Yes} \\
		\hline
		\hspace{0.6cm}\color{blue}{${\cal{O}}(m^{-(2\ell+1)})$} & HD Maxwell & {\rm Yes} & {\rm Yes} & {\rm Yes} \\
		\hline
	\end{tabular}
	\caption{Comparison among the photon's effective action terms induced by the spinor $\rm QED_3$, scalar $\rm QED_3$ and $\rm RS$-$\rm QED_3$.}
	\label{tab1}
\end{table}

\subsection{The case for $\omega=4$}

In this section, we want to explore the renormalizability of the RS model in detail for the case of $\omega = 4$, because the expression \eqref{eq8} is divergent in this case, and this divergence has a structure of an HD term.
Moreover, HD terms are known to change the renormalizability properties of the model.
Therefore, this proposal wishes to highlight aspects of the type of counter-terms that are necessary to handle the divergences.

In this case, the factor $\Gamma\left(2-\frac{\omega}{2}\right)$ in \eqref{eq8} leads to the divergence.
Hence, we apply the dimensional regularization to compute \eqref{eq8}, giving us
\begin{equation} \label{eq12}
i\Pi^{\mu\nu}\left(p\right)=\frac{i\tilde{g}^{2}}{72\pi^{2}m^4}\int_{0}^{1}dx~ N^{\mu\nu}(x) ~ \Gamma\left(\frac{\varepsilon}{2}\right) \left(\frac{4\pi\mu^{2}}{ x (x-1)p^{2}+m^{2}}\right)^{\frac{\varepsilon}{2}},
\end{equation}
with the parameter $\varepsilon = 4-\omega$.
The energy scale $\mu$ is introduced to make a dimensionless coupling as $g^2 \to \tilde{g} ^2 \mu^{4-\omega}$ and also  $N^{\mu\nu}(x)$ refers to the tensorial part of the integrand in \eqref{eq8}, which is finite.
Finally, the regularization yields as usual
\begin{equation}
\Gamma\left(\frac{\varepsilon}{2}\right) \left(\frac{4\pi\mu^{2}}{ x (x-1)p^{2}+m^{2}}\right)^{\frac{\varepsilon}{2}} \simeq \frac{2}{\varepsilon}-\ln\Big(\frac{ x (x-1)p^{2}+m^{2}}{4\pi\mu^2}\Big)-\gamma_{_E},
\end{equation}
where $\gamma_{_E}$ is the Euler-Mascheroni constant.

Hence, using the above property and integrating over the Feynman parameter at $\varepsilon \rightarrow 0$, the divergent part of the amplitude \eqref{eq12} reads
\begin{equation} \label{eq13}
i\Pi^{\mu\nu}_{\rm div.}(p) =-\frac{i g^2}{54 \pi^2  }\frac{1}{\varepsilon} \frac{p^2}{m^2}\Big(\frac{p^2}{m^2}-4\Big)
\left(p^2 \eta^{\mu\nu}-p^{\mu}p^{\nu}\right).
\end{equation}
Here, we observe that the divergent term \eqref{eq13} is not like as in the usual QED$_4$ (a renormalizable theory) and this jeopardizes the renormalizability of the RS-QED$_4$, because this divergence requires the presence of higher-derivative (Lee-Wick) counter-terms not present in the original dynamics.
This feature resembles the non-renormalizability of the Hilbert-Einstein action and the need to introduce higher-order curvature (higher-derivative) terms to achieve a consistent renormalization of gravity \cite{Stelle:1976gc}.

The presence of such bad UV behavior of the polarization tensor \eqref{eq13} in RS-QED$_4$ can be understood by the fact that the spin-$3/2$ mode cannot improve renormalizability (in comparison with QED$_4$) since the UV asymptotic behavior of the RS propagator \eqref{eq4} is worse than of the usual Dirac propagator.

As a conclusion of this section, it would be nice to compare the one-loop photon polarization tensor arising from coupling to the massive particles with different spins in the table \ref{tab2}.

There, in table \ref{tab2}, we can observe that the tensor structure is the same between the QED and RS-QED, for all three cases, changing only its numerical coefficients due to the different spin content.
Moreover, as mentioned above, for the cases $\omega=3,4$ the RS-QED includes the HD terms in the parity even sector (proportional to $Q^{\mu\nu}$) when compared with QED and SQED, leading to the need of higher-derivative (Lee-Wick) counter-terms to renormalize the divergence in four-dimensional RS-QED.

\begin{table}[h!]
	\centering
\begin{tabular}{|c|c|c|c|}
	 		\hline
	 		& & & \\
			\color{blue}{Dimension} & \color{blue}{$i\Pi^{\mu\nu}_{_{\rm SQED}}(p)$}
			 &
			\color{blue}{$i\Pi^{\mu\nu}_{_{\rm QED}}(p)$}
			 & \color{blue}{$i\Pi^{\mu\nu}_{_{\rm RS-QED}}(p)$}\\
				 & & & \\
								\hline
								 & & & \\
			{\color{blue}$\omega=2$} &\quad\quad$-\frac{ig^2}{12\pi m^2}Q^{\mu\nu}$~~~\quad& $-\frac{ig^2}{6\pi m^2}Q^{\mu\nu}$ & $-\frac{2ig^2}{\pi m^2}Q^{\mu\nu}$\\
		 & & & \\
		\hline
		 & & & \\
		{\color{blue}$\omega=3$} &\quad $-\frac{ig^2}{24\pi m}Q^{\mu\nu}$ & $-\frac{ig^2}{12\pi m}Q^{\mu\nu}+\frac{g^2}{4\pi}T^{\mu\nu}$ ~&  $\frac{ig^2}{30\pi m^3}p^2Q^{\mu\nu}+\frac{3g^2}{4\pi}T^{\mu\nu}$ \\
			 & & & \\
			\hline
		 & & & \\
	{\color{blue}$\omega=4$} & \quad $-\frac{ig^2}{24\pi^2 \varepsilon}Q^{\mu\nu}$ & $	-\frac{ig^2}{6\pi^2 \varepsilon}Q^{\mu\nu}$ & $-\frac{i g^2}{54 \pi^2 \varepsilon } \frac{p^2}{m^2}\left(\frac{p^2}{m^2}-4\right)Q^{\mu\nu}$ \\
		 & & & \\
		\hline
		\end{tabular}
	\caption{Comparison among the one-loop photon polarization tensor obtained by the massive scalar $\rm QED$, spinor $\rm QED$ and $\rm RS$-$\rm QED$ in various dimensions. For simplicity in notation, we introduced two tensors $Q^{\mu\nu}\equiv p^2\eta^{\mu\nu}-p^{\mu}p^{\nu}$ and $T^{\mu\nu}\equiv\epsilon^{\mu\nu\alpha}p_{\alpha}$.}
	\label{tab2}
\end{table}

Regarding the numerical coefficients of $Q^{\mu\nu}$ in the table \ref{tab2}, we notice that
	\begin{equation}		
		\Pi^{\mu\nu}_{_{\rm QED_2}}=2\Pi^{\mu\nu}_{_{\rm SQED_2}},~~ \Pi^{\mu\nu}_{_{\rm QED_3}}\Big|_{\rm P-even}=2\Pi^{\mu\nu}_{_{\rm SQED_3}}, ~~\Pi^{\mu\nu}_{_{\rm QED_4}}=4\Pi^{\mu\nu}_{_{\rm SQED_4}}.
	\end{equation}
	This behavior originates from the spin property which is included in gamma matrices. The amplitude of the polarization tensor in spinor QED is proportional to the trace of four gamma matrices so that $\omega=2,3$ yields the coefficient 2 and $\omega=4$ yields the coefficient 4.
Since the special form of Feynman vertex and propagator in RS-QED produces a complicated amplitude (which yields HD terms for  $\omega=3,4$), we do not have such a clear comparison to SQED and QED.

\section{Dynamical mass generation for the gauge field}
\label{sec4}
Once we have determined the expressions for the polarization tensor at $\omega=2,3$ we can finally analyze how the gauge field (dynamical) mass behaves in the RS-QED.
We can examine the dynamical mass generation in terms of the physical pole in the complete propagator.
With this aim, we now proceed to obtain a general expression for the gauge field propagator, including the radiative corrections.

In general, due to its gauge invariance, the polarization tensor can be cast as
\begin{equation} \label{tensor-PT}
\Pi^{\mu\nu}(p)\Big|_{ loop- level}=\Big(\eta^{\mu\nu}-\frac{p^{\mu}p^{\nu}}{p^{2}}\Big)\Pi_{\rm e}
+i\epsilon^{\mu\nu\alpha}p_{\alpha}\Pi_{\rm o},
\end{equation}
in which $\Pi_{\textrm{e}}$ and $\Pi_{\textrm{o}}$ are the 1PI form factors related to the parity even and odd sectors of the theory.
Naturally, there can be more of these form factors if the thermal effects are considered \cite{Weldon:1996kb}, and also the geometric factors (e.g. noncommutativity of spacetime \cite{Brandt:2001ud,Ghasemkhani:2015tqu}).

The above structure \eqref{tensor-PT} allows us to obtain the following general expression for the complete propagator as
\begin{equation} \label{eq20}
i\mathcal{D}^{\mu\nu}(p)=\frac{p^{2}-\Pi_{\textrm{e}}}{\Delta_{\rm phys}}\Big(\eta^{\mu\nu}-\frac{p^{\mu}p^{\nu}}{p^{2}}\Big)+\frac{\xi}{p^{2}}\frac{p^{\mu}p^{\nu}}{p^{2}}+\frac{\Pi_{\textrm{o}}}{\Delta_{\rm phys} }i\epsilon^{\mu\nu\alpha}p_{\alpha}.
\end{equation}
Here, $\xi$ is the gauge parameter. The physical pole defined in terms of $\Delta_{\rm phys}=\left(p^{2}-\Pi_{\textrm{e}}\right)^2-p^{2}\Pi_{\textrm{o}}^{2}$ is the main quantity of interest in order to examine the dynamical mass generation for the gauge field.

We shall now examine explicitly the physical pole $\Delta_{\rm phys}$ for the $\omega=2,3$ cases as well as discuss some implications.
Moreover, whenever pertinent, we will establish a comparison between the obtained results for the RS-QED model with the usual QED, so that we can trace the modifications due to the RS fields.

\subsection{One-loop photon propagator in $\omega=2$}

In the $\omega= 2$ case, we observe from the RS-QED expression \eqref{PT_W2} that the form factor is $\Pi_{\rm e}(m)=- \frac{2g^{2} p^2}{\pi m^{2}}$.
Hence, the one-loop propagator \eqref{eq20} at Landau gauge ($\xi=0$) takes the form
\begin{align}
i\mathcal{D}^{\mu\nu}(p)\Big|_{one-loop}= \frac{1}{p^{2}\big(1+\frac{2g^{2}}{\pi m^{2}}\big)} \Big(\eta^{\mu\nu}-\frac{p^{\mu}p^{\nu}}{p^{2}}\Big),
\end{align}
which means that the physical pole reads $\Delta_{\rm phys}=p^{2}\big(1+\frac{2g^{2}}{\pi m^{2}}\big)$.
Therefore, the propagator is only corrected by a coefficient at the one-loop level as it happens in the massive Schwinger model \cite{Coleman:1975pw}.
Here, we notice that since the mass dimension of the coupling constant in two dimensions is $[g]=1$ so the quantity $(\frac{g}{m})$ is a dimensionless coupling.

On the other hand, for the case of massless RS fields, we notice from \eqref{PT_W2_mass} that the form factor is $\Pi_{\rm e}(0)=\frac{g^2}{\pi }$, which leads to the propagator $ i\mathcal{D} ^{\mu\nu}\left(p\right) = \frac{\eta^{\mu\nu}}{p^{2} - m_\gamma^2 }$, in which we observe the presence of a gauge invariant mass $m_\gamma^2 = g^2 /\pi $.
This is precisely the famous Schwinger result $m_{\rm schw}^2 = g^2 /\pi $ \cite{Schwinger:1962tp}.

In summary, from our previous results, we conclude that the RS model in $\omega=2$ behaves physically the same as the Schwinger model (in both massive and massless phases).

Another point worth to remark is that in the usual Schwinger model, all radiative contributions to the polarization tensor beyond the one-loop vanish identically, which renders the Schwinger mass to be exact.
It would be interesting to study this higher-loop behavior in the (massless) RS model and determine whether the higher-order corrections to the photon mass are still absent.

\subsection{One-loop photon propagator in $\omega=3$}

Now we proceed and study the effects of the HD terms into the gauge field propagator in $\omega=3$.
For the RS-QED in $\omega=3$ we can compute the parity even part from \eqref{eq10}
\begin{equation}
\Pi_{\rm e}\left(p\right) =  \frac{g^2}{30\pi m^3}~p^4 ,
\end{equation}
while the parity odd part is obtained from  \eqref{eq10b}
\begin{equation}
\Pi_{\rm o}\left(p\right)= \frac{3g^2}{4\pi} -\frac{g^2}{6\pi m^2}~p^{2} .
\end{equation}
In this case, we have the one-loop propagator \eqref{eq20} at Landau gauge ($\xi=0$) as below
\begin{equation} \label{prop_3}
i\mathcal{D} ^{\mu\nu} \left(p^{2}\right) =  \frac{p^2\left(1- \lambda p^2\right)\left(\eta^{\mu\nu}-\frac{p^{\mu}p^{\nu}}{p^{2}}\right)+i(\kappa-\omega p^2)\varepsilon^{\mu\nu\alpha}p_{\alpha} }{p^2\big[p^2(1-\lambda p^2)^2-(\kappa-\omega p^2)^2 \big]},
\end{equation}
in which we have introduced the coefficients $\lambda$, $\kappa$ and $\omega$ such as
\begin{equation}
\lambda=\frac{g^2}{30\pi m^{3}},\quad \kappa=\frac{3g^2}{4\pi},\quad\omega=\frac{g^2}{6\pi m^2}.
\end{equation}
They will assist us in performing some limits and discuss cases of interest.
In addition, a point to remark is that since the mass dimension of the coupling constant $g$ in $\omega=3$ is $[g]=1/2$, then we conclude that $[\lambda]=-2$, $[\kappa]=1$ and $[\omega]=-1$ and hence the quantity $\lambda p^2$ in \eqref{prop_3} is a dimensionless quantity.

The physical pole in \eqref{prop_3} reads $\Delta_{\rm phys}=p^2\big[p^2(1-\lambda p^2)^2-(\kappa-\omega p^2)^2 \big]$. Let us then consider some implications of this expression.
The first point is that when $\lambda=\omega=\kappa=0$ we have the usual Maxwell propagator in \eqref{prop_3}.

Moreover, in the absence of the HD terms $ (\lambda=\omega=0)$, we have
\begin{equation} \label{prop_4}
i D^{\mu\nu}= \frac{p^2\eta^{\mu\nu}-p^{\mu}p^{\nu}+i\kappa\varepsilon^{\mu\nu\alpha}p_{\alpha}}{p^2(p^2-\kappa^2)},
\end{equation}
which is exactly the propagator of the Maxwell-Chern-Simons theory \cite{Dunne:1998qy}.
One can see in \eqref{prop_4} that there is a non-zero pole at $p^2=\kappa^2$ that generates a mass term for the photon as $m_{\gamma}^2=\frac{3g^2}{4\pi}$ \cite{Dunne:1998qy}.

On the other hand, in the presence of the HD terms, we can study the physical poles of the photon propagator by solving the following equation
\begin{equation}
\Delta_{\rm phys}=p^2\big[p^2(1-\lambda p^2)^2-(\kappa-\omega p^2)^2 \big]=0,
\end{equation}
which is a cubic polynomial in $\chi =p^2$.
There are three cases of interest to examine the non-trivial poles $p^2  \neq 0$:

\begin{itemize}

	\item In the presence of HD-Chern-Simons term $\omega \neq 0$ and the absence of HD-Maxwell term $\lambda=0$, we find two real solutions as follows
	\begin{equation}
		\chi_1=\frac{1+2\omega \kappa-\sqrt{1+4\omega\kappa}}{2\omega^2}
		,\quad 	\chi_2=\frac{1+2\omega \kappa+\sqrt{1+4\omega\kappa}}{2\omega^2}.
	\end{equation}
It is easy to see that the value of $\chi_1$ is positive so we have two different positive poles.
	\item Moreover, in the presence of HD-Maxwell term $\lambda \neq 0$ and the absence of the HD-Chern-Simons term $\omega=0$, we have one real solution
\begin{equation}
\chi=\frac{2}{3\lambda}+\frac{2^{1/3}}{3\Sigma^{1/3}}+\frac{\Sigma^{1/3}}{2^{1/3}3\lambda^{2}},
\end{equation}		
in which we have defined $\Sigma=27\kappa^{2}\lambda^{4}-2\lambda^{3}+3\sqrt{3}\kappa\lambda^{3}\sqrt{27\kappa^{2}\lambda^{2}-4\lambda}$.

\item At last, we have the general case, in the presence of both HD-Maxwell term $\lambda \neq 0$ and HD-Chern-Simons term $\omega \neq 0$, we obtain only one real solution as follows
		
 \begin{equation} \label{root}
\chi=\frac{2\lambda+\omega^{2}}{3\lambda^{2}}+\frac{2^{1/3}\left(\lambda^{2}-6\kappa\lambda^{2}\omega +4\lambda\omega^{2}+\omega^{4}\right)}{3\lambda^{2}\Omega^{1/3}}+\frac{\Omega^{1/3}}{  2^{1/3}3 \lambda^{2}},		
		 \end{equation}
where we have introduced by simplicity of notation the following quantities
\begin{align}
\Omega=&\,\Gamma+\sqrt{\Theta} , \\
\Gamma =&\, 27\kappa^{2}\lambda^{4}-2\lambda^{3}-36\kappa\lambda^{3}\omega+15\lambda^{2}\omega^{2}-18\kappa\lambda^{2}\omega^{3}+12\lambda\omega^{4}+2\omega^{6}, \\
\Theta=&\,4\left(6\kappa\lambda^{2}\omega-\lambda^{2}-4\lambda\omega^{2}-\omega^{4}\right)^{3} +\Gamma ^{2}.
     \end{align}		
Actually, by a simple check in Mathematica, one observes that the root \eqref{root} is always positive and corresponds to a physical pole. 		
  \end{itemize}

\section{Final remarks}
\label{sec5}

In this manuscript, we have discussed the dynamical mass generation for the photon field in the Rarita-Schwinger QED.
We wanted to address the generation of the Schwinger mass in $\omega =2$ and also the (topological) Chern-Simons mass in $\omega =3$.
The results were very interesting, with an emphasis on the cases of $\omega =3,4$, in which the presence of higher-order (derivative) terms changed the physical behavior of the models.

Since our main interest was to examine the physical pole of the (one-loop) propagator, to address the mass generation, the main objects to be evaluated were the form factors (parity even and odd pieces) related to the polarization tensor in the cases of $\omega =2,3,4$.
With this purpose,  we established the $\omega$-dimensional (massive and massless) propagator and vertex function, which are used to determine the polarization tensor in $\omega$ dimensions.

In the two-dimensional case, the RS-QED yields similar results as the usual QED: the massive RS-QED reproduces the physical behavior of the massive Schwinger model; on the other hand, the massless RS-QED gave precisely the same result as the Schwinger model, generating the gauge invariant photon mass $m_\gamma^2 = g^2 /\pi $. It is worth studying the presence/absence of higher-order corrections to the photon mass in massless RS-QED$_{2}$.

Moreover, in the three-dimensional case, the RS-QED has a significantly different behavior than QED in the parity even sector, while the parity odd has the same structure.
In the parity even sector, we observe the absence of the Maxwell term (in the low-energy expansion), and the leading contributions are due to HD terms, but since they are finite the renormalizability is preserved.
On the other hand, in the parity odd sector the RS-QED yields the usual Chern-Simons term in the leading order.
The physical poles were discussed in three cases: i) only the HD-Chern-Simons term, which gives two real roots, ii) only the HD-Maxwell term, which gives only one real solution, iii) the general case, with both HD-Maxwell and HD-Chern-Simons terms, which also has one real solution.

Due to the presence of HD terms in the polarization tensor at $\omega=3$, we sought to analyze the amplitude for $\omega=4$ to grasp a better understanding of the HD terms in the renormalizability.
Actually, the RS-QED$_4$ also presents a UV divergence \eqref{eq13}, but it is not like as in the usual QED$_4$.
This structure jeopardizes the renormalizability of RS-QED$_4$, because it requires the presence of higher-derivative (Lee-Wick) counter-terms.
This result can be understood from the fact that the UV asymptotic behavior of the RS propagator \eqref{eq4} is evidently worse than of the usual Dirac propagator.

\section*{Acknowledgments}

The authors would like to thank the anonymous referee for his/her comments to improve this paper. We are also grateful to M.M. Sheikh-Jabbari
for discussions on related topics. R.B. acknowledges partial support from Conselho Nacional de Desenvolvimento Cient\'ifico e Tecnol\'ogico (CNPq Project No. 306769/2022-0).


\global\long\def\link#1#2{\href{http://eudml.org/#1}{#2}}
 \global\long\def\doi#1#2{\href{http://dx.doi.org/#1}{#2}}
 \global\long\def\arXiv#1#2{\href{http://arxiv.org/abs/#1}{arXiv:#1 [#2]}}
 \global\long\def\arXivOld#1{\href{http://arxiv.org/abs/#1}{arXiv:#1}}


\end{document}